\documentclass[aps, prl, amsmath,amssymb,notitlepage, onecolumn, nofootinbib]{revtex4-1}

\usepackage[utf8]{inputenc} % set input encoding (not needed with XeLaTeX)

\usepackage{graphicx}
\graphicspath{{Figures-reply/}}
\usepackage{amsmath}
\usepackage[margin=0.9in]{geometry}
\usepackage{epstopdf}
\usepackage{booktabs}
\usepackage{color}
\begin{document}

\title{Real-time interferometric quantum random number generation on chip}
\author{Thomas Roger$^1$, Taofiq Paraiso$^1$, Innocenzo De Marco$^{1,2}$, Davide G. Marangon$^1$, Zhiliang Yuan$^1$, Andrew J. Shields$^1$}
\affiliation{$^1$ Toshiba Research Europe Ltd., 208 Cambridge Science Park, Cambridge, UK\\
$^2$ School of Electronic and Electrical Engineering, University of Leeds, Leeds, LS2 9JT, UK}
\begin{abstract}
We demonstrate on-chip quantum random number generation at high data rates using the random phases of gain-switched laser pulses. Interference of the gain-switched pulses produced by two independent semiconductor lasers is performed on a photonic integrated circuit (PIC) and the resulting pulse train is received and processed in real-time using homebuilt capture electronics consisting a field programmable gate array (FPGA) and a 10-bit digitizer. Random numbers with low correlation coefficient are shown for pulse clock rates of 1 GHz and data rates of 8 Gbps. The random numbers are also shown to successfully pass all tests within the National Institute for Standards and Technology (NIST) test suite. The system provides genuine random numbers in a compact platform that can be readily integrated into existing quantum cryptographic technology. \\
\end{abstract}

\maketitle

\noindent \textit{Introduction-} Random numbers are an essential resource in many modern applications such as quantum key distribution (QKD) \cite{Uchida, Kanter, Argyris, Williams, Wayne, Xu, Gabriel}. As demand grows for quantum transmitters with higher data rates and smaller footprint, the random number generators we use must also meet these challenges \cite{Dynes}. Specifically, they should produce high quality random numbers at gigahertz clock rates. Indeed any predictability in the numbers would lead to a significant decrease in the security of the cryptographic technique and low data rates could cause a bottleneck to the transmitted data speed. Quantum physics has been touted as the ultimate route to produce true random numbers due to the intrinsic stochastic nature of quantum phenomena. It is now becoming clear that true random number generators for use in the most demanding applications will rely on quantum methods \cite{Kofler, Hall, Yuan1}.\\

Quantum random number generators (QRNGs) have been demonstrated based on the detection of single photon events \cite{Yan}, vacuum fluctuations \cite{Gabriel, Shen, Symul}, phase noise \cite{Li, Qi, Jofre, Yuan2, Xu2, Abellan} and quantum nolocality \cite{Pironio, Yang}. A particular approach that allows for very high data rate random number generators has been demonstrated using gain-switching of a distributed feedback laser (DFB) \cite{Yuan2}. In this approach a laser is operated above threshold and periodically brought below threshold by modulation with a radio frequency (RF) signal. By depleting the laser cavity on each cycle of the RF signal the lasing process is continuously seeded by the quantum vacuum. The result is to produce short pulses of light that are phase independent. This is due to the inherent uncertainty on the phase of the quantum vacuum  \cite{Williams}. Current implementations of this technique have been made using bulk optics and laboratory electronics, as such they are large in size and costly \cite{Sun}. For successful large scale deployment of QKD systems it is desirable to considerably reduce this requirement.\\

Photonic integrated circuit (PIC) technology allows for scalable optical devices that can be readily incorporated into existing infrastructure. For instance, PICs are already widely used for coherent classical communications \cite{Goldsmith}. Advances in PIC technology utilising indium phosphide (InP) have allowed for on-chip photodetectors, lasers, couplers and multiplexers to be integrated in a single compact platform or device. These advances have stimulated  the use of PIC technology for quantum optics applications. There are now a number of examples of on-chip QKD, QRNG and single photon detection in the literature \cite{Francis, Rafaelli, Sibson}. Many examples of phase-noise based QRNGs use a single pulsed laser and an asymmetric Mach-Zehnder interferometer (AMZI) to interfere consecutive pulses. Here we use two independent lasers in order to help reduce correlations between adjacent pulses. This simple design, which uses two lasers that interfere at a 50:50 beamsplitter has recently been demonstrated using bulk optics \cite{Sun}. \\
 
As PIC technology matures and its implementation to cryptographic devices increases we will see a need for low cost and scalable random number generation. Existing on-chip QRNGs tend to rely on active elements off chip and suffer from low clock rates \cite{Francis, Rafaelli}. However, a notable example exploiting the interference of a continuous-wave laser with a pulsed laser, producing random numbers at 1 Gb/s, has been presented recently \cite{Abellan2}. Here, we demostrate a dual pulsed laser QRNG PIC producing high quality random numbers at up to 8 Gbps. The on-chip implementation of the device provides a platform for the stable production of random numbers in real-time. We show on-chip interference of the gain-switched pulses at a clock rate of 1 GHz and test the resulting numbers using the NIST test suite. The random numbers are shown to have low correlation coefficient across the pulse train. \\

\begin{figure*}[ht!]
\centering{
\includegraphics[scale = 0.5]{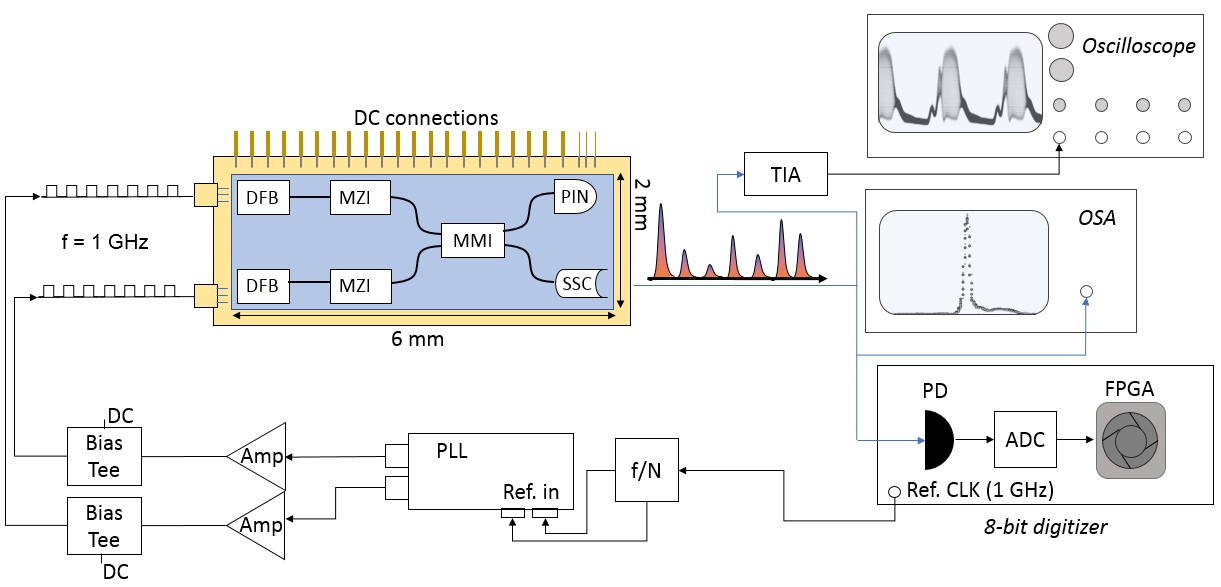}
\caption{Experimental schematic- the QRNG PIC consists of dual distributed feedback lasers (DFB), two tuneable Mach-Zehnder interferometer switches (MZI), a 50:50 2x2 coupler (MMI), fast photodiode (PIN) and an output spot size converter (SSC). The DFB lasers are driven via RF tracks and the wavelength tuned via heaters with direct current (DC) connections at the top edge of the PIC. The output gain-switched pulses from the DFB lasers pass through tuneable switches in order to balance the pulse powers, which are tunable by applying a DC voltage to heaters on each arm of interferometer arms. The balanced laser pulses are interfered at the MMI and coupled out via the SSC. The PIC sits inside a RF package providing connections to the RF and DC inputs. The RF signal to the lasers is generated by a phase locked loop (PLL) referenced to the data capture board (1 GHz clock) via a frequency down converter. The outputs of the two-channel PLL are amplified (Vpp 3.44 V) and combined with a DC bias (1.25 V) to operate the DFBs in gain-switching mode. The output of QRNG PIC is coupled via a lensed tapered fibre and sent to the data capture board, operating at 8-bit resolution. The pulse train is monitored via optical spectrum analyser (OSA) and fast oscilloscope for characterisation of the device.}
\label{fig:setup}}
\end{figure*}

\noindent \textit{Experiment-} Figure~\ref{fig:setup} shows the experimental scheme. RF waveforms are produced using a two channel phase-locked loop (PLL) and amplified to achieve the desired voltage swing. The two channels of the PLL serve to drive the two semiconductor lasers onboard the PIC. The output of each of the lasers passes through a Mach-Zehnder interferometer, which is used to control the transmitted power. The laser outputs are combined onto a 2 x 2 50:50 multi-mode inteferometer (MMI) and subsequently coupled out from the PIC by an angled spot size converter (SSC). The angle is chosen to minimise reflections from the output facet. The second output of the MMI is coupled to a PD that is used to monitor the lasers output. The photonic circuit that acts as the QRNG is housed within a RF package and wire bonded to the package for access to the RF and direct current (DC) connections to the lasers, heaters and photodiode (PD) on the PIC. The RF package sits on an aluminium heat sink and a bespoke printed circuit board (PCB) is used to address the DC connections along the side of the package. The RF signals are input via micro-miniature coaxial (MMCX) connectors on the left hand edge of the package. The PIC itself consists of the two independent DFB lasers, which are driven by ground-signal-ground electrical inputs situated on the left hand edge of the PIC. A lensed-tapered fibre is used to collect the light from the SSC and couples the output from the RF package to be measured via homebuilt capture electronics which include a photodiode, 10-bit analog-to-digital converter (ADC) and FPGA for processing of the random bit string. Processing of the raw 10-bit data via the FPGA reduces the output stream to 8 bits. A 1 GHz clock that is synchronised to the capture electronics is used to provide a differential reference to the PLL via a 5-bit frequency down-converter. The PIC was fabricated using Indium Phosphide (InP) with a standard integration process \cite{InP}. The InP platform allows for the active elements (lasers) to be integrated on chip rather than coupled in externally.  \\

A schematic of the QRNG-PIC is shown in figure~\ref{fig:setup}. The PIC, housed within the RF package, sits on top of a thermoelectric cooling peltier and the temperature is monitored via an negative temperature coefficient (NTC) thermistor. The temperature was maintained at 22$\pm 0.001 ^{\circ}$C. The temperature stability minimises any additional correlations to the phase of the pulses that may adversely affect the randomness of the interference. The DFB lasers were driven well below saturation with a bias of 44 mA (1.25 V) and an RF amplitude of 3.44 Vpp, coupled via a bias-tee. An output power of 35 $\mu$W for each laser was measured at the output of the RF package. Fine adjustments to the output power were made by applying a DC voltage to the heaters on the MZI of each laser in order to accurately match the power of each laser. The laser power was set at the start of the measurement and no significant change ($<1\%$) in the output power was measured during the course of the experiment. The DC voltages for the lasers and MZIs were supplied using a multi-channel source measure unit. The beating signal was detected by the FPGA capture electronics and monitored using a high speed trans-impedance amplifier (TIA) and fast oscilloscope. The wavelength of each laser was monitored using a high resolution optical spectrum analyser.\\

The random numbers are extracted from the interference of the two laser pulses at a 50:50 beamsplitter via a photodetector. Following the example of bulk optics phase-noise based QRNGs  \cite{Xu2, Zhou, Sun} we may write the measured signal as $V(t) \propto E_1E_2\cos[\Delta\omega_0t+\theta_1-\theta_2]$, where $E_i$ and $\theta_i$ are the electric field amplitude and phase of the two lasers $i = 1,2$ and $\Delta\omega_0 = \omega_1-\omega_2$ is the beat frequency. As noted previously, when operated in the gain-switching regime the phase of each laser pulse is random due to seeding from the vacuum, resulting in a random output level $V(t)$. This interference provides the basis of the quantum random signal, however, in addition to this signal there are also contributions to the noise of a classical nature. These arise due to the inherent noise of the photodetector, analog-to-digital converter and FPGA processing electronics. Although we are predominantly showcasing the PIC and capture board electronics speed to produce high output rate random numbers, in the discussion we show that if a more stringent approach is taken, accounting for all sources of randomness it would be possible to output fully secure random numbers with a slightly reduced data rate.\\

\noindent \textit{Results-} In the following section we show the results of the experiment, which is depicted in figure 1. We drive the DFB lasers with identical 1 GHz square RF waveforms, which are temporally matched. The resultant pulses are combined on chip at the 50:50 2x2 coupler and interfere to produce the random intensity level pulse train, which is converted to the random number bit string at the digitizer. \\

\begin{figure*}[ht!]
\centering\includegraphics[scale = 0.35]{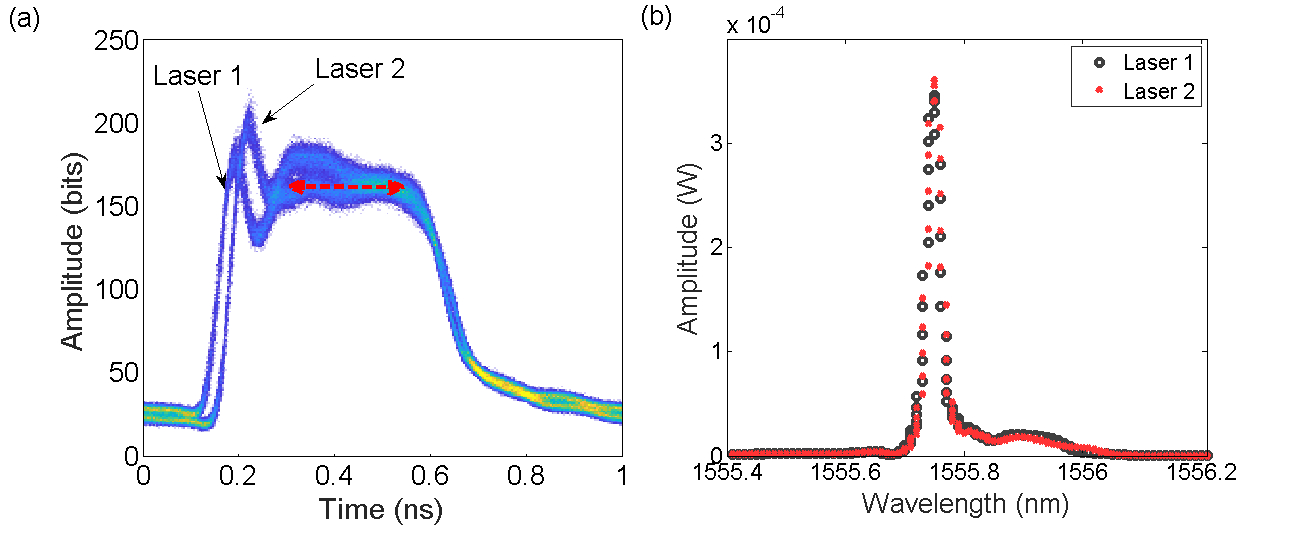}
\caption{(a) Optical output of the gain-switched DFB lasers when operating in the steady-state condition (marked with red dashed arrow). An initial overshoot is followed by 200 ps of equilibrium where the carrier density within the laser cavity is balanced between depletion via stimulated emission and excitation from the driving signal. (b) The spectra of the DFB lasers are shown. The spectra of lasers is matched by detunning each of the DFB heaters.}
\label{fig:spectral}
\end{figure*}

Gain-switched laser diodes are now routinely used to generate phase incoherent short light pulses \cite{Pataca, Gobby}. However, these short pulses do not lend themselves to interferometric RNG due to pulse jitter and frequency chirp, thus requiring extremely precise temporal overlap of the pulses. The timing jitter and chirp are a result of the short build up and depletion of carriers within the cavity, thus causing rapid changes to the cavity optical constants. These problems can be overcome by driving the lasers at a higher DC bias. In this case an initial overshoot to the pulse shape relaxes to a steady state, at which point the chirp is minimised, see section marked by red arrow in figure \ref{fig:spectral}~(a). It should be noted that during the initial overshoot the pulse is strongly chirped and therefore we do not measure the interference at this point in the experiment. The equilibrium position persists for approximately 200 ps and permits a stable platform for interference of the pulses. A signature of the lasers operating within this  regime is a sharp peaked spectrum, illustrating that almost no frequency chirp exists. The spectra for both lasers are shown in figure~\ref{fig:spectral} (b). \\

\begin{figure*}[ht!]
\centering{
\includegraphics[scale = 0.33]{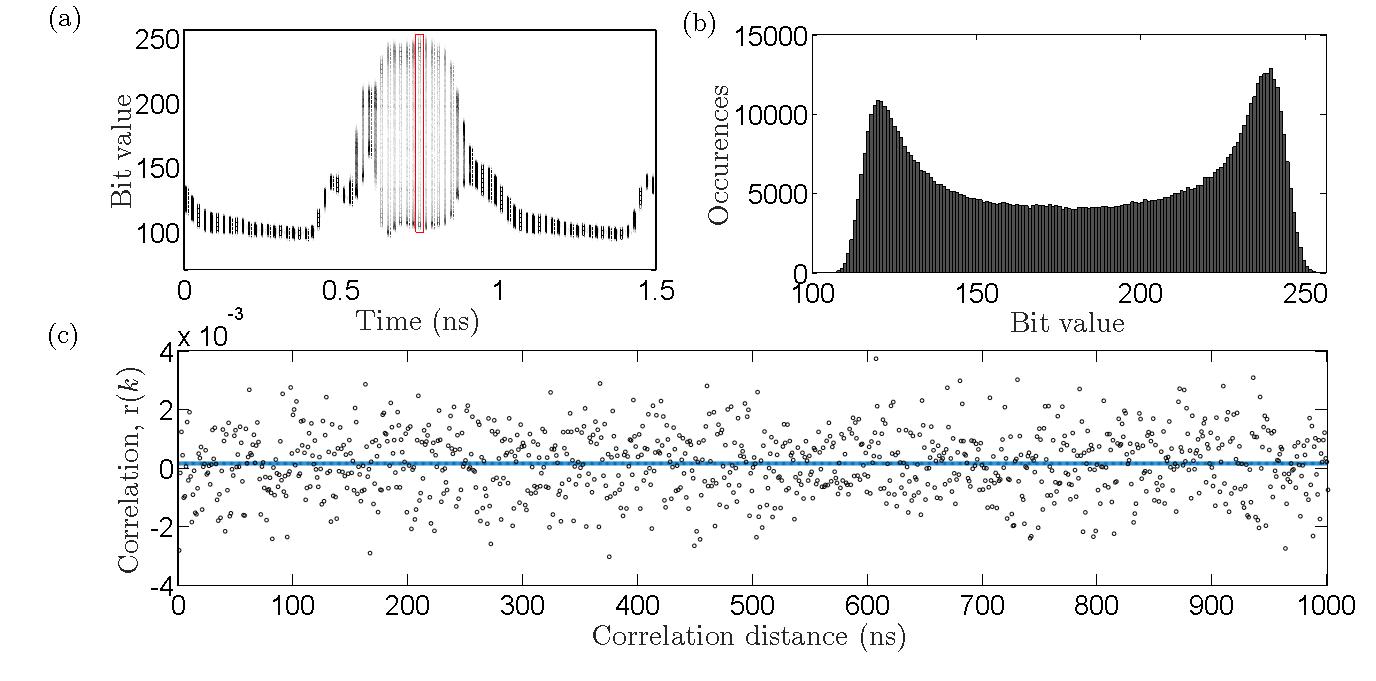}
\caption{(a) The interference of the two DFB lasers is measured using a transimpedance amplifier (TIA) and fast oscilloscope in persistent display mode. Each vertical line corresponds to one sample of the oscilloscope, which we operated at 50 GSps ($\delta\tau = 20$ ps). (b) Histogram showing the pulse intensity distribution measured at the center of the pulse interference marked by the red rectangle in (a). (c) The autocorrelation trace of the pulse train sampled at interval $k$, which was equal to the pulse repetition rate i.e. 1 GHz. There is no correlation present across all the pulses. The mean of the correlation is depicted by the solid blue line and was measured to be $1.5\times10^{-4}$.}
\label{fig:histcorr}}
\end{figure*}

 In figure~\ref{fig:histcorr} (a) we show the high-visibility interference of the two lasers operating within the steady-state regime. Driving the lasers in the steady state condition provides a characteristic double-peaked interference distribution. In figure \ref{fig:histcorr} (b), we show the distribution of detected interference events at the 8-bit digitizer. The low bit values correspond to destructive interference while high values to constructive interference measured at a specific time sample within the centre of the pulse. The double peaked structure is a result of identical phase evolution of each pulse within the train when there is negligible chirp as has been shown in previous studies \cite{Yuan2}. Data from the central portion of the pulse ($\sim$0.7 ns) corresponds to the histogram data in (a).\\

In figure \ref{fig:histcorr} (c), we show the autocorrelation of the pulse train output from the QRNG PIC. The autocorrelation function for the pulse events, $x_i$, is calculated using $r(k)= \sum_{i=1}^{N-k} (x_i-{\bf x})(x_{i+k}-{\bf x})/\sum_{i=1}^N(x_i-{\bf x})^2$, where $k$ is the pulse separation and the autocorrelation is performed over the entire sequence of N = 20000 samples and displayed in figure \ref{fig:histcorr} (c) over the adjacent 1000 samples.  We note that in this data, when driving the lasers at 1 GHz, we find that all values over the entire correlation distance show no bias outside those expected due to statistical fluctuations, indicating that the QRNG is producing true random numbers and can be used at the highest data output rates. The full experimental run was performed over 860 $\mu$s and shows no correlation outside those shown in figure \ref{fig:histcorr}(c).\\

\noindent \textit{Discussion}-  As discussed previously, the digitized interference events occur with a probability that follows an arcsine ``double peaked" distribution. This implies that the ADC outputs cannot be used directly as random numbers, because they are not uniformly distributed. Therefore, one could often guess the output of the generator to be values corresponding to constructive or destructive interference.
In order to remove this bias, a \textsl{finite impulse response} (FIR) filter processing is applied to the ``raw random samples''. This processing, introduced in \cite{Yuan2}, scrambles the numbers such that they span the whole 8-bit range $\left\{0,255\right\}$ evenly. If $x(n)$ denotes a raw sample input to the FIR filter, the output $y(n)$ is given by $y(n) = b_0 x(n)+b_1 x(n-1)+\dots + b_M x(n-M) \mod 2^8$, where $b_i=M!/(i!(M-i)!)$ are binomial coefficients and we set $M=7$, for this proof of principle. It is worth stressing that this operation is performed in real time by the FPGA. Although the FIR filter does not provide the same security level as a randomness extractor, it is an effective debiasing technique that does not affect the generation rate and makes it possible to pass the statistical tests of randomness \cite{Marangon}. Other filters used for de-biasing of the raw data have been proposed in the literature \cite{Kanter1, Kanter2}, these are based on converting the raw data into a time series consisting of the derivative of the ADC signal amplitude. These methods can also be implemented in real time. For higher levels of security, i.e. for randomness generation secure in a quantum information framework one may envision post-processing of the data using Trevisan's \cite{De} or Toeplitz \cite{Ma} extractors. These methods are outside the scope of this work but their implementation to future QRNG devices is the focus of further study. However, we may still estimate the performance of our device for extraction of fully secure random numbers. To do so we use the model outlined in \cite{Sun} to estimate the min-entropy $H_{\text{min}}(X|E)$, corresponding to the maximal amount of unpredictable bits $X$ that the extractor can distil per 8-bit sample conditioned on the side information $E$. According to the model, $E$ represents the classical side information i.e. the information available to a potential attacker. We estimate $H_{\text{min}}(X|E)=4.09$ bits per 8-bit sample. This value represents the content of unpredictable random bits that can be distilled on average per sample once a strong randomness extractor is applied. With efficient implementation onto the FPGA capture board, the secure generation rate would then be $>$4 Gb/s.  \\

In Table~\ref{nist_results} we show the result of applying the NIST test suite, which runs 17 tests on the output random numbers. For each of the tests we supplied 1000 strings of random bits, with each string containing 1 million bits. The p-value is quoted for each of the tests and the proportion of the bits that passed the test is given. Each of the 17 tests was passed.\\

\begin{table}
\centering
\begin{tabular}{l c c c}    \hline
\emph{Statistical test} & \emph{P value} &  \emph{Proportion} &  \emph{Result} \\\hline
Frequency                                & 0.9061 & 0.989 & Success \\
Block Frequency                      &0.0835 & 0.992 & Success \\
Cumulative Sums                     &0.8817 & 0.986 & Success \\
Cumulative Sums                     &0.3838 & 0.989 & Success \\
Runs                                         &0.8580 & 0.986 & Success \\
Longest Run                             &0.2968 & 0.992 & Success \\
Rank                                         &0.6371 & 0.986 & Success \\
FFT                                           &0.2812 & 0.993 & Success \\
Non Overlapping Template       & 0.5045 & 0.990 & Success \\
Overlapping Template              &0.8343 & 0.983 & Success\\
Universal                                  &0.1238 & 0.987 & Success\\
Approximate Entropy              &0.3330 & 0.990 & Success\\
Random Excursions                 & 0.4151 & 0.989 & Success\\
Random Excursions Variant    & 0.4882 & 0.992 & Success\\
Serial                                        &0.2012 & 0.990 & Success\\
Serial                                        &0.4101 & 0.995 & Success\\
Linear Complexity                   &0.3361 & 0.992 & Success\\\bottomrule
\hline
\end{tabular}
\caption{Results of the NIST test battery applied on $10^3$ strings, each having a length of $10^6$ bits.}\label{nist_results}
\end{table}

\noindent \textit{Summary-} We have presented a high speed photonic integrated QRNG based on the InP platform. The chip is driven by bespoke electronics and data is captured and analysed using a homebuilt FPGA digitizer. The PIC utilises a dual DFB laser scheme whereby the lasers are driven in the gain-switching regime and combined on chip to produce a random intensity level pulse train. The PIC and electronics system is capable of producing high quality (low correlation) random numbers at 8 Gbps, which are output in real-time due to the high-speed platform (FPGA) of the capture electronics. The small form factor of the optical device lends itself to packaging in standard optical modules and future prototypes will look to embed the PIC, driving and capture electronics within a single PCB. The device can be readily integrated with complementary metal oxide semiconductor (CMOS) electronics, allowing for deployment in commercial communications systems or into existing cryptographic technology. \cite{Gottesman, Lo}. \\

\acknowledgements{This work has been partially funded by the Innovate UK project EQUIP, as part of the UK National Quantum Technologies Programme.  IDM and DGM acknowledge funding from the European Union’s Horizon 2020 research and innovation programme under the Marie Skłodowska-Curie grant agreement No 675662 and 750602, respectively.  }

\end{document}